\documentclass[twoside]{article}
\usepackage{fleqn,espcrc2}
\usepackage{graphicx}

\title{High Energy Neutrinos from Gamma-Ray Bursts
}

\author{Eli Waxman
\thanks{Incumbent of the Beracha foundation career development chair}
\thanks{Work supported in part 
by BSF Grant 9800343, AEC Grant 38/99 and MINERVA Grant.}
\address{Dept. of Condensed Matter Physics, Weizmann Institute of Science,
Rehovot 76100, Israel}}

\begin{document}

\begin{abstract}
 
Observations suggest that $\gamma$-ray bursts (GRBs) 
are produced by the dissipation 
of the kinetic energy of a relativistic fireball. In this talk, 
recent work on the production of high energy neutrinos by GRB fireballs is
reviewed.
A significant fraction of GRB energy is expected to be converted to
an accompanying burst of high energy neutrinos.
Photomeson interactions produce a burst of $\sim100$~TeV 
neutrinos in coincidence with the GRB, and a burst of 
$\sim10^{18}$~eV neutrinos following the GRB on a time scale of 10~s. 
Inelastic $p$--$n$ nuclear collisions result in the
production of a burst of $\sim10$~GeV neutrinos in coincidence with the GRB.
Planned 1~km$^3$ neutrino telescopes 
are expected to detect tens of 100~TeV
neutrino events, and several $10^{18}$~eV
events, correlated with GRBs per year. A suitably densely spaced detector
may allow the detection of several 10~GeV events per year.
The detection of high-energy neutrino events correlated 
with GRBs will allow to constrain GRB progenitor models and 
to test the suggestion that GRBs accelerate protons to $\sim10^{20}$~eV.
Moreover, such detection will allow to test for neutrino properties,
e.g. flavor oscillations (for which upward moving $\tau$'s would be 
a unique signature) and coupling
to gravity, with an accuracy many orders 
of magnitude better than is currently possible.

\end{abstract} 

\maketitle

\section{Introduction}

The widely accepted interpretation of the
phenomenology of GRBs,
bursts of 0.1 MeV--1 MeV photons lasting for a few seconds
\cite{Fishman}, is that the
observable effects are due to the dissipation of the kinetic energy
of a relativistically expanding wind, a 
``fireball,'' whose primal cause is not yet known
(see \cite{fireballs} for review). 
During the past three years, delayed low energy emission 
(afterglow) of GRBs has been discovered in X-ray, 
optical, and radio wave bands (see \cite{AG_ex_review} for review).
Afterglow observations confirmed the cosmological origin of the bursts,
through the redshift determination of several GRB host-galaxies,
and confirmed standard model predictions of afterglows
that result from the collision of an expanding fireball with
its surrounding medium (see \cite{AG_th_review} for review). 

The physical conditions in the fireball dissipation region 
allow Fermi shock acceleration of protons
to energy $>10^{20}{\rm eV}$ \cite{W95a,VMU95}.  
The average rate at which energy is emitted as $\gamma$-rays
by GRBs is  
comparable to the energy generation rate of $>10^{19}$~eV
cosmic-rays in a model where
such ultra-high energy cosmic-rays (UHECRs) 
are produced by a cosmological distribution of sources 
\cite{W95a,W95b}. These two facts suggest that GRBs and UHECRs have
a common origin (see \cite{My_revs} for review). 

Emission of high energy neutrinos 
is expected to accompany the observed GRB photons, 
due to photomeson interactions of high energy protons with
fireball photons \cite{WnB9799,AG_nus},
and due to proton-neutron
inelastic collisions \cite{Derishev99,BnM00,MnR_nus}. In this review, 
recent work on the production of high energy neutrinos by GRB fireballs is
summarized, and the implications for planned $\sim1{\rm km}^3$ 
neutrino telescopes (the ICECUBE extension of AMANDA,
ANTARES, NESTOR; see \cite{Halzen99} for review) are discussed.
The fireball model is briefly described in \S2. Proton acceleration
in GRB fireballs is discussed in \S3. 
High energy neutrino production in fireballs and its implications for future
high energy neutrino detectors are discussed in \S4.

\section{The fireball model}

In the fireball model of GRBs
\cite{fireball86}, a compact source, of linear scale
$r_0\sim10^7$~cm, produces a wind characterized by an average luminosity 
$L\sim10^{52}{\rm erg\,s}^{-1}$ and mass loss rate $\dot M=L/\eta c^2$.
At small radius, 
the wind bulk Lorentz factor, $\Gamma$, 
grows linearly with radius, until most of the wind energy is converted
to kinetic energy and $\Gamma$ saturates at $\Gamma\sim\eta\sim300$.
Variability of the source on time scale $\Delta t$, resulting
in fluctuations in the wind bulk Lorentz factor $\Gamma$ on similar
time scale, then leads to internal shocks \cite{internal}
in the expanding fireball at a radius
$r_i\approx\Gamma^2c\Delta t$.
If the Lorentz factor variability within the wind is significant,
internal shocks reconvert a substantial 
part of the kinetic energy to internal energy. 
It is assumed that this energy is then radiated as 
$\gamma$-rays by synchrotron and inverse-Compton emission of
shock-accelerated electrons.

In this model, the observed
$\gamma$-ray variability time, $\sim r_i/\Gamma^2 c\approx\Delta t$,
reflects the variability time of the underlying source, and the GRB
duration, $T\sim10$s, 
reflects the duration over which energy is emitted from the
source. A large fraction of bursts detected by BATSE show variability
on the shortest resolved time scale, $\sim10$~ms \cite{Woods95}, and some show
variability on shorter time scales, $\sim1$~ms \cite{Bhat92}.
This sets the constraint on underlying source size, 
$r_0<c\Delta t\sim10^7$~cm. The wind must be expanding relativistically, 
with a Lorentz factor $\Gamma\sim300$, 
in order that
the fireball pair-production optical depth be small for observed 
high energy, $\sim100$~MeV, GRB photons \cite{Gamma}.

As the fireball expands, it drives a relativistic shock (blast-wave)
into the surrounding gas. At early time, $t\ll T$,
the fireball is little affected by this external interaction. 
At late time, $t\gg T$,
most of the fireball energy is transferred to the 
surrounding gas, and the flow approaches the Blandford-McKee
self-similar flow \cite{BnM76}.
The shock driven into the ambient medium continuously heats fresh gas, and
accelerates relativistic electrons which produce through synchrotron emission 
the delayed radiation, ``afterglow,'' 
observed on time scales of days to months. 
As the shock-wave decelerates, due to accumulation of ambient gas
mass, the emission shifts with time to
lower frequencies. For expansion into uniform density gas, the
shock Lorentz factor decreases with time
according to $\Gamma_{BM}(t)\propto t^{-3/8}$.

During the transition to self-similar expansion, which 
takes place on (observed)
time scale comparable to $T$, reverse shocks propagate into the fireball
ejecta and decelerate it. At this stage the shocked plasma
expands with the self-similar Lorentz factor, which 
for expansion into uniform density gas is given by (e.g. \cite{AG_nus})
\begin{equation}
\Gamma_{BM}(t=T)\simeq 245\left({E_{53}\over n_0}\right)^{1/8}T_1^{-3/8},
\label{eq:Gamma}
\end{equation}
while the unshocked fireball
ejecta propagate with the original expansion Lorentz factor, 
$\Gamma=\eta>\Gamma_{BM}(t=T)$. Here,
$E=10^{53}E_{53}$~erg is the total fireball energy,
$T=10 \, T_1$~s, and $n=1 \, n_0{\rm cm}^{-3}$ is the ambient gas density.
$n_0=1$ is typical to the interstellar medium. 

Internal shocks are expected 
to be ``mildly'' relativistic in the fireball 
comoving frame, i.e. characterized by Lorentz factor 
$\Gamma_i-1\sim1$ in the wind rest frame, 
since adjacent shells within the wind are expected to
expand with Lorentz factors which do not differ by more than an
order of magnitude. Moreover, the reverse shocks are also expected
to be mildly relativistic, since the ratio $\eta/\Gamma_{BM}(t=T)$ is
not far from unity.

\section{UHECRs from GRB fireballs}

\subsection{Fermi acceleration in GRBs}

In the fireball model, the observed GRB and afterglow radiation is produced
by synchrotron emission of shock accelerated
electrons. In the region where electrons are accelerated, 
protons are also expected to be
shock accelerated. This is similar to what is thought to occur in supernovae 
remnant shocks \cite{Bland87}. We consider below proton acceleration 
in internal and reverse fireball shocks.
Since these shocks are mildly relativistic,
we expect results related to particle
acceleration in sub-relativistic shocks (see, e.g.,
\cite{Bland87} for review) to be valid for the present
scenario. In particular, the predicted energy distribution of accelerated
protons is $dN_p/d\epsilon_p\propto \epsilon_p^{-2}$.

Two constraints must be satisfied by
fireball wind parameters in order to allow proton acceleration to
$\epsilon_p>10^{20}$~eV in internal shocks \cite{W95a}:
\begin{equation}
\xi_B/\xi_e>0.02\Gamma_{2.5}^2 \epsilon_{p,20}^2L_{\gamma,52}^{-1},
\label{eq:xi_B}
\end{equation}
and
\begin{equation}
\Gamma>130 \epsilon_{p,20}^{3/4}\Delta t^{-1/4}_{-2}.
\label{eq:G_min}
\end{equation}
Here, $\epsilon_p=10^{20}\epsilon_{p,20}$~eV, 
$\Delta t=10^{-2}\Delta t_{-2}$~s,
$\Gamma=10^{2.5}\Gamma_{2.5}$ is the plasma expansion Lorentz factor,
$L_{\gamma}=10^{52}L_{\gamma,52}{\rm erg/s}$
is the $\gamma$-ray luminosity, $\xi_B$ is the 
fraction of the wind energy density which is carried by magnetic field,
$4\pi r^2 c\Gamma^2 (B^2/8\pi)=\xi_B L$, 
and $\xi_e$ is the fraction of wind energy carried by shock
accelerated electrons. To obtain the constraints for acceleration in 
reverse shocks, $\Delta t$ should be replaced by $T$.
Since the electron energy is lost radiatively,
$L_\gamma\approx\xi_e L$. The first constraint must be satisfied in order 
for the proton acceleration time $t_a$ to be smaller than the wind expansion 
time. The second constraint must be satisfied in order for the 
synchrotron energy loss time of the proton to be larger than $t_a$.

The constraints that must be satisfied to allow acceleration of protons 
to energy $>10^{20}$~eV are remarkably similar to those inferred from
$\gamma$-ray observations. $\Gamma>100$ is implied by observed
$\gamma$-ray spectra, and 
magnetic field close to equipartition, $\xi_B\sim1$, is required
in order for electron synchrotron emission to account for the observed
radiation. 

We have assumed in the discussion so far that the fireball is spherically 
symmetric. However, since a jet-like fireball behaves as if it were
a conical section of a spherical fireball as long as the jet opening
angle is larger than $\Gamma^{-1}$, our
results apply also for a jet-like fireball 
(we are interested only in processes that occur when
the wind is ultra-relativistic, $\Gamma\sim300$, prior to 
significant fireball deceleration). For a jet-like wind, $L$ in our
equations should be understood as the luminosity the fireball
would have carried had it been spherically symmetric.

\subsection{Energy generation rate}

The observed GRB redshift distribution implies a typical GRB 
$\gamma$-ray energy
of $E\approx10^{53}$~erg, and a GRB rate of 
$R_{\rm GRB}\sim3/{\rm Gpc}^3{\rm yr}$ at $z\sim1$. 
The present, $z=0$, rate is less well constrained, since most observed 
GRBs originate at redshifts $1\le z\le2.5$ \cite{AG_ex_review}. 
Present data
are consistent with both no evolution of GRB rate with redshift, and 
with strong evolution (following, e.g.,
star formation rate), in which $R_{\rm GRB}(z=1)/R_{\rm GRB}(z=0)\sim10$
\cite{GRB_z}.
The energy observed in $\gamma$-rays reflect the fireball
energy in accelerated electrons. If 
accelerated electrons and protons carry similar energy 
(as indicated by afterglow observations \cite{Freedman})
then the GRB cosmic-ray production rate is
\begin{figure}
\includegraphics[width=3in]{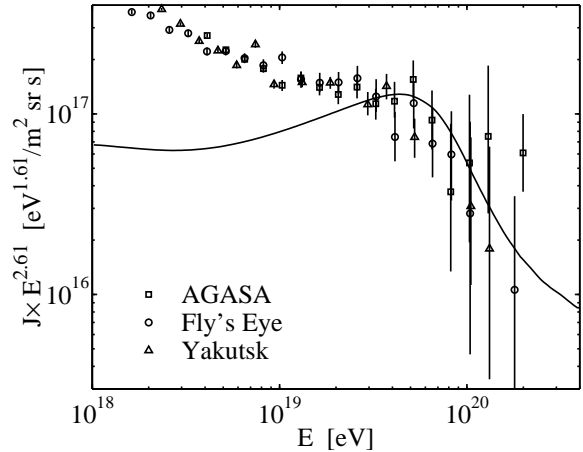}
\caption{
The UHECR flux expected in a cosmological model, where high-energy protons 
are produced at a rate $(\epsilon_p^2 d\dot n_p/d\epsilon_p)_{z=0}=
0.8\times10^{44}{\rm erg/Mpc}^3{\rm yr}$ as predicted in the GRB model 
[Eq. (\ref{eq:cr_rate})], compared to the Fly's Eye, Yakutsk and AGASA data. 
$1\sigma$ flux error bars are shown. The highest energy points are derived
assuming the detected events (1 for Fly's Eye and Yakutsk, 
4 for AGASA) represent a
uniform flux over the energy range $10^{20}$~eV--$3\times10^{20}$~eV.}
\label{fig1}
\end{figure}
\begin{equation}
\epsilon_p^2 (d\dot n_p/d\epsilon_p)_{z=0}= 
10^{44}\zeta {\rm erg/Mpc}^3{\rm yr},
\label{eq:cr_rate}
\end{equation}
with $\zeta$ in the range of $\sim10^{-0.5}$ to $\sim10^{0.5}$.
This result is consistent with more detailed analyses \cite{MnM98plus},
where $R_{\rm GRB}(z=0)\times E$ in the range of 0.2 to 2 times 
$10^{44} {\rm erg/Mpc}^3{\rm yr}$ is obtained (under the assumption
that $R_{\rm GRB}$ is proportional to star formation rate)
for different models of star formation rate evolution.

In Fig. 1 we compare the UHECR spectrum,
reported by the Fly's Eye \cite{Bird934}, the Yakutsk \cite{Yakutsk}, 
and the AGASA \cite{Takeda98} experiments, 
with that predicted by the GRB model (assuming GRB rate follows
star formation rate). 
The generation rate (\ref{eq:cr_rate}) of high energy protons
is remarkably similar to that 
required to account for the flux of $>10^{19}$~eV cosmic-rays
(The flux at lower energies is most likely dominated by heavy nuclei
of Galactic origin \cite{Bird934}, as indicated by 
the flattening of the spectrum at $\approx10^{19}$~eV and by the evidence
for a change in composition at this energy \cite{composition}).

The suppression of model flux above $10^{19.7}$~eV is  
due to energy loss of high energy protons
in interaction with the microwave background, i.e. to the ``GZK cutoff''
\cite{GZK}. The available data do not allow to determine
the existence (or absence) of the ``cutoff''
with high confidence. The AGASA results show an excess (at a
$\sim2.5\sigma$ confidence level) of events compared to model
predictions above $10^{20}{\rm eV}$. This excess is not confirmed,
however, by the other experiments. Moreover, since the $10^{20}{\rm eV}$
flux is dominated by sources at distances $<100\ {\rm Mpc}$, over
which the distribution of known astrophysical systems
(e.g. galaxies, clusters of galaxies) is inhomogeneous,
 significant deviations from model predictions presented
in Fig. 1 for a uniform source distribution are expected at this energy
\cite{W95b}.
Clustering of cosmic-ray sources leads
to a standard deviation, $\sigma$, in the expected number, $N$, of 
events above $10^{20}$ eV, given by 
$\sigma /N = 0.9(d_0/10 {\rm Mpc})^{0.9}$
\cite{WnB_clustering}, where $d_0$ is the unknown scale
length of the source correlation function and $d_0\sim10$ Mpc 
for field galaxies.

Although the rate of GRBs out to a distance of 100~Mpc from
Earth, the maximum distance traveled by $>10^{20}$~eV protons, is
in the range of $10^{-2}$ to $10^{-3}{\rm yr}^{-1}$, the number of 
different GRBs contributing to the flux of $>10^{20}$~eV protons at any given
time may be large. This is due to the dispersion $\tau$ in proton
arrival time, which is expected due to deflection by
inter-galactic magnetic fields and may 
be as large as $10^{5}$~yr, implying that
the number of sources contributing to the 
flux at any given time may be as large as 
$\tau\times10^{-2}{\rm yr}^{-1} =10^3$ \cite{W95b}.

\section{High energy Neutrinos}

\subsection{Photomeson production}

\subsubsection{Internal shocks neutrinos, $\sim10^{14}$~eV}

The decay of charged
pions, $\pi^+\rightarrow\mu^++\nu_\mu
\rightarrow e^++\nu_e+\overline\nu_\mu+\nu_\mu$, created in interactions
between fireball photons and
accelerated protons, 
results in the production of high energy neutrinos \cite{WnB9799}. 
The key relation
is between the observed photon energy, $\epsilon_\gamma$,
and the accelerated proton's energy, $\epsilon_p$,
at the photomeson threshold  of the $\Delta$-resonance.
In the observer frame,
\begin{equation}
\epsilon_\gamma \, \epsilon_{p} = 0.2 \, {\rm GeV^2} \, \Gamma^2.
\label{eq:keyrelation}
\end{equation}
For $\Gamma\approx300$ and $\epsilon_\gamma=1$~MeV,
the typical observed energy of photons emitted in internal shocks,
we see that characteristic proton energies
$\sim 10^{16}$~eV
are required to produce neutrinos from pion decay.
Typically, the neutrinos receive $\sim 5$\% of
the proton energy, leading to neutrinos of $\sim 10^{14}$~eV.

The fraction $f_\pi(\epsilon_p)$
of proton energy lost to pion production is given by the ratio
of the wind expansion time, $\Gamma\Delta t$ in the wind frame
at the internal shock stage,
and the proton photopion energy loss time. 
For observed GRB photon spectra, 
$dN_\gamma/d\epsilon_\gamma\propto \epsilon_\gamma^{-\beta}$ 
with $\beta\simeq1$ for photon energies below the break energy
$\epsilon_\gamma^b\approx1$~MeV 
and $\beta\simeq2$ for $\epsilon_\gamma>\epsilon_\gamma^b$, 
$f_\pi$ is independent of energy for $\epsilon_p>0.2\Gamma^2{\rm GeV}^2/
\epsilon_\gamma^b\sim10^{16}$~eV,
\begin{equation}
f_\pi\approx0.2{L_{\gamma,52}\over
(\epsilon_\gamma^b/1{\rm MeV})\Gamma_{2.5}^4 \Delta t_{-2}},
\label{fpi}
\end{equation}
and $f_\pi\propto\epsilon_p$ for $\epsilon_p<10^{16}$~eV.
For a typical burst, $E\approx10^{53}$~erg at redshift $z\sim1$, 
Eq. (\ref{fpi})
implies a neutrino fluence of \cite{WnB9799}
\begin{equation}
\epsilon_\nu^2\Phi_{\nu_x}\approx
10^{-3}\left({f_\pi\over0.2}\right)\left({\epsilon_\nu\over10^{14}
{\rm eV}}\right)^{\alpha} {\rm GeV\over cm^2},
\label{JGRB}
\end{equation}
where $\alpha=0$ for $\epsilon_\nu>10^{14}{\rm eV}$
and $\alpha=1$ for $\epsilon_\nu<10^{14}{\rm eV}$, and
$\nu_x$ stands for $\nu_\mu$, $\bar\nu_\mu$ and $\nu_e$.

The neutrino spectrum (\ref{JGRB}) is
modified at high energy, where neutrinos are produced by the decay
of muons and pions whose life time 
exceeds the characteristic time for
energy loss due to adiabatic expansion and synchrotron emission 
\cite{RnM98,WnB9799}.
The synchrotron loss time is determined by the energy density of the
magnetic field in the wind rest frame.
For the characteristic parameters of a GRB wind, 
synchrotron losses are the dominant effect, leading to strong suppression of
$\nu$ flux above
$\sim10^{16}$~eV.

\subsubsection{Afterglow neutrinos, $\sim10^{18}$~eV}

During the transition to self-similarity, 
protons and electrons are accelerated to high energy 
in the reverse shocks. High energy protons 
may interact with the 10~eV--1~keV photons 
radiated by the electrons, 
to produce through pion decay a burst of duration $\sim T$ 
of ultra-high energy, $10^{17}$--$10^{19}$~eV, neutrinos as indicated by
Eq. (\ref{eq:keyrelation}) \cite{AG_nus}.

While afterglows have been detected in several cases, reverse
shock emission has only been identified for GRB 990123 \cite{Akerlof99}.
Both the detections and the non-detections are consistent with shocks
occurring with typical model parameters \cite{SPMR_0123},
suggesting that reverse shock emission may be common.
The predicted neutrino emission depends, however, upon parameters
of the surrounding medium that can only be estimated once
more observations of the prompt optical afterglow emission are available.

If the density of gas surrounding the fireball is 
$n\sim1{\rm cm}^{-3}$,
a value typical to the inter-stellar medium and consistent with
GRB 990123 observations, then the synchrotron emission of reverse
shock electrons is expected to peak in the X-ray band, 
$\epsilon_\gamma^b\approx1$~keV, with luminosity $L_X\approx2\times
10^{50}{\rm erg/s}$
\cite{AG_nus}. Using these parameters in Eq. (\ref{fpi}), 
replacing $\Delta t$ with $T\approx10$~s and recalling that $\Gamma=250$
at the reverse shock stage (see \S2), we find 
$f_\pi\approx0.01$ for protons of energy $10^{19}$~eV. Thus, the expected 
neutrino fluence for a typical burst, $E\approx10^{53}$~erg at $z\sim1$, is
\cite{AG_nus}
\begin{equation}
\epsilon_\nu^2\Phi_{\nu_x}\approx
10^{-4.5}\left({\epsilon_\nu\over10^{17}
{\rm eV}}\right)^\alpha {\rm GeV\over cm^2},
\label{JAG}
\end{equation}
where $\alpha=1/2$ for $\epsilon_\nu>10^{17}{\rm eV}$
and $\alpha=1$ for $\epsilon_\nu<10^{17}{\rm eV}$.
Here too, $\nu_x$ stands for $\nu_\mu$, $\bar\nu_\mu$ and $\nu_e$.
The value of $\alpha$, $\alpha=1/2$ for $\epsilon_\nu>10^{17}{\rm eV}$, 
corresponding to photomeson interactions of photons with energy
below the break $\epsilon_\gamma^b\approx1$~keV, differs from the value
$\alpha=0$ in the case of internal shocks discussed in \S4.1.1, since
the low energy, $\epsilon_\gamma<\epsilon_\gamma^b$,
photon spectrum is different in the two cases:
$dN_\gamma/d\epsilon_\gamma\propto \epsilon_\gamma^{-\beta}$ 
with $\beta=1$ for internal shocks, and $\beta=3/2$ 
for reverse shocks.

Some GRBs may result from the collapse of a massive star
(e.g. \cite{Woosley_Chevalier99}), in
which case the fireball is expected to expand into a pre-existing wind. 
For typical wind parameters, the
transition to self-similar behavior takes place at a radius where the
wind density is $n\approx10^4{\rm cm}^{-3}\gg 1{\rm cm}^{-3}$. The higher
density implies a lower Lorenz factor of the expanding plasma during
the transition stage, and hence a larger fraction of proton energy lost
to pion production. Protons of energy
$\epsilon_p\ge 10^{18}$~eV lose all their energy to pion production
in this case, and a typical GRB at $z\sim1$
is expected to produce a neutrino fluence  \cite{AG_nus,Dai00}
\begin{equation}
\epsilon_\nu^2\Phi_{\nu_x}\approx
10^{-2.5}\left({\epsilon_\nu\over10^{17}
{\rm eV}}\right)^\alpha {\rm GeV\over cm^2},
\label{JAGw}
\end{equation}
where $\alpha=0$ for $\epsilon_\nu>10^{17}{\rm eV}$
and $\alpha=1$ for $\epsilon_\nu<10^{17}{\rm eV}$.

The neutrino flux is expected to be strongly suppressed at energy
$>10^{19}$~eV, since protons are not expected to be
accelerated to energy $\gg10^{20}$~eV.

\subsection{Inelastic $p$-$n$ collisions}

The acceleration, $\Gamma\propto r$, of fireball plasma emitted from the
source of radius $r_0$ (see \S2) is driven by radiation pressure. Fireball
protons are accelerated through their coupling to the electrons, which
are coupled to fireball photons. Fireball neutrons, which are expected to
exist in most progenitor scenarios, are coupled to protons by nuclear
scattering as long as the comoving $p$-$n$ scattering time
is shorter than the comoving wind expansion time $r/\Gamma c=r_0/c$. 
As the fireball plasma expands and accelerates, the proton density decreases,
$n_p\propto r^{-2}\Gamma^{-1}$, and neutrons may become decoupled.
For $\eta>\eta_{pn}$, where
\begin{equation}
\eta_{pn}\approx400L_{52}^{1/4}r_{0,7}^{-1/4}
\label{eq:Gamma_pn}
\end{equation}
and $r_0=10^7r_{0,7}$~cm, neutrons decouple
from the accelerating plasma prior to saturation, $\Gamma=\eta$,
at $\Gamma=\eta_{pn}^{4/3}\eta^{-1/3}$ \cite{BnM00}. In this
case, relativistic relative velocities between protons and neutrons arise,
which lead to pion production through inelastic nuclear collisions. Since
decoupling occurs at a radius where the collision time is similar to wind
expansion time, each $n$ leads on average to one pair of $\nu\bar\nu$.
The typical comoving neutrino energy, $\sim50$~MeV, implies an observed
energy $\sim10$~GeV. A typical burst, $E=10^{53}$~erg at $z=1$, with
significant neutron to proton ratio and $\eta>400$ will therefore
produce a fluence $F(\nu_e+\bar\nu_e)\sim0.5F(\nu_\mu+\bar\nu_\mu)\sim
10^{-4}{\rm cm}^{-2}$ of $\sim10$~GeV neutrinos.

Relativistic relative $p$-$n$ velocities, leading to neutrino
production through inelastic collisions, may also result from diffusion
of neutrons between regions of the fireball wind with large difference in
$\Gamma$ \cite{MnR_nus}. 
If, for example, plasma expanding with very high Lorentz factor,
$\Gamma>100$, is confined to a narrow jet surrounded by a slower, 
$\Gamma\sim10$ wind, internal collisions within the slower wind can heat
neutrons to relativistic temperature, leading to significant diffusion
of neutrons from the slower wind into the faster jet. Such process may
operate for winds with $\eta<400$ as well as for $\eta>400$, and may
lead, for certain (reasonable) wind parameter values, to $\sim10$~GeV
neutrino flux similar to that due to $p$-$n$ decoupling in a 
$\eta>400$ wind.

\subsection{Implications}

The predicted fluence of $\sim10^{14}$~eV neutrinos produced
by photomeson interactions in internal fireball shocks, Eq. (\ref{JGRB}),  
implies a detection probability $\sim10^{-1.5}$ per burst
in planned $1{\rm km}^3$ neutrino telescopes \cite{Halzen99}, corresponding
to detection of several tens of muon induced neutrinos 
per year correlated in time and direction 
with GRBs, given the observed GRB rate of 
$\approx10^3{\rm yr}^{-1}$. The predicted fluence of 
$\sim10^{17}$~eV neutrinos, produced
by photomeson interactions during the onset of fireball interaction 
with its surrounding medium, Eqs. (\ref{JAG},\ref{JAGw}), 
implies a detection probability $\sim10^{-4.5}$ in the case of fireball 
expansion into typical inter-stellar medium gas, and $\sim10^{-2.5}$ 
in the case of fireball expansion into a pre-existing massive star wind.
Several muon induced neutrinos per year in a $1{\rm km}^3$ 
detector are expected in the latter case. 

Detection of high energy neutrinos
will test the shock acceleration mechanism and the suggestion that
GRBs are the sources of ultra-high energy protons, since $\ge10^{14}$~eV
($\ge10^{18}$~eV)
neutrino production requires protons of energy $\ge10^{16}$~eV
($\ge10^{19}$~eV). The dependence of $\sim10^{14}$~eV neutrino flux
on wind Lorentz factor, Eq. (\ref{fpi}), and the dependence of 
$\sim10^{17}$~eV neutrino flux
on fireball environment, imply that the detection of high energy
neutrinos will also provide constraints on wind Lorentz factor
\cite{HnH00}, and on the GRB progenitor.

Inelastic $p$-$n$ collisions may produce $\sim10$~GeV neutrinos
with a fluence of $\sim10^{-4}{\rm cm}^{-2}$ per burst, 
due to either $p$-$n$ decoupling in a wind 
with high neutron fraction and high, $>400$, 
Lorentz factor \cite{Derishev99,BnM00}, 
or to neutron diffusion in a wind with, e.g., 
strong deviation from spherical symmetry \cite{MnR_nus}.
The predicted number of events in a $1{\rm km}^3$ neutrino telescope
is $\sim10{\rm yr}^{-1}$. Such events may be detectable in a suitably
densely spaced detector. Detection of $\sim10$~GeV neutrinos will 
constrain the fireball neutron fraction, and hence the GRB progenitor.

Detection of neutrinos from GRBs could be used to
test the simultaneity of
neutrino and photon arrival to an accuracy of $\sim1{\rm\ s}$
($\sim1{\rm\ ms}$ for short bursts), checking the assumption of 
special relativity
that photons and neutrinos have the same limiting speed.
These observations would also test the weak
equivalence principle, according to which photons and neutrinos should
suffer the same time delay as they pass through a gravitational potential.
With $1{\rm\ s}$ accuracy, a burst at $100{\rm\ Mpc}$ would reveal
a fractional difference in limiting speed 
of $10^{-16}$, and a fractional difference in gravitational time delay 
of order $10^{-6}$ (considering the Galactic potential alone).
Previous applications of these ideas to supernova 1987A 
(see \cite{John_book} for review), where simultaneity could be checked
only to an accuracy of order several hours, yielded much weaker upper
limits: of order $10^{-8}$ and $10^{-2}$ for fractional differences in the 
limiting speed and time delay respectively.

The model discussed above predicts the production of high energy
muon and electron neutrinos. 
However, if the atmospheric neutrino anomaly has the explanation it is
usually given, oscillation to $\nu_\tau$'s with mass $\sim0.1{\rm\ eV}$
\cite{atmo}, then
one should detect equal numbers of $\nu_\mu$'s and $\nu_\tau$'s. 
Up-going $\tau$'s, rather than $\mu$'s, would be a
distinctive signature of such oscillations. 
Since $\nu_\tau$'s are not expected to be produced in the fireball, looking
for $\tau$'s would be an ``appearance experiment.''
To allow flavor change, the difference in squared neutrino masses, 
$\Delta m^2$, should exceed a minimum value
proportional to the ratio of source
distance and neutrino energy \cite{John_book}. A burst at $100{\rm\ Mpc}$ 
producing $10^{14}{\rm eV}$ neutrinos can test for $\Delta m^2\ge10^{-16}
{\rm eV}^2$, 5 orders of magnitude more sensitive than solar neutrinos.


\begin{thebibliography}{12}

\bibitem{Fishman} 
  Fishman, G. J. \& Meegan, C. A., ARA\&A {\bf 33}, 415 (1995).
\bibitem{fireballs} 
  Piran, T., in {\it Unsolved Problems In Astrophysics}, eds.
  J. N. Bahcall and J. P. Ostriker, 343-377 (Princeton, 1996).
\bibitem{AG_ex_review}
  Kulkarni, S. R. {\it et al.}, 
  To appear in Proc. of the 5th Huntsville Gamma-Ray Burst Symposium
  (astro-ph/0002168)
\bibitem{AG_th_review}
  M\'esz\'aros, P., A\&AS {\bf 138}, 533 (1999).
\bibitem{W95a} 
  Waxman, E.,  Phys. Rev. Lett. {\bf 75}, 386 (1995).
\bibitem{VMU95} 
  Vietri, M., Ap. J. {\bf 453}, 883 (1995); 
  Milgrom, M. \& Usov, V., Ap. J. {\bf 449}, L37 (1995).
\bibitem{W95b} 
  Waxman, E.,  Ap. J. {\bf 452}, L1 (1995).
\bibitem{My_revs}
  Waxman, E., Physica Scripta T85, 117 (2000); Waxman, E., Nucl. Phys.
  B (Proc. Suppl.) {\bf 87}, 345 (2000).
\bibitem{WnB9799} 
  Waxman, E., \& Bahcall, J. N., Phys. Rev. Lett. {\bf 78}, 2292 (1997);
  Waxman, E., \& Bahcall, J. N., Phys. Rev. {\bf D59}, 023002 (1999).
\bibitem{AG_nus}
  Waxman, E., \& Bahcall, J. N., Ap. J., in press (hep-ph/9909286).
\bibitem{Derishev99}
  Derishev, E. V., Kocharovsky, V. V., \& Kocharovsky, Vl. V,
  Ap. J. {\bf 521}, 640 (1999).
\bibitem{BnM00}
  Bahcall, J. N., \& M\'esz\'aros, P., Phys. Rev. Lett., in press 
  (hep-ph/0004019).
\bibitem{MnR_nus}
  M\'esz\'aros, P., \& Rees, M., Ap. J., submitted (astro-ph/0007102).
\bibitem{Halzen99}
  Halzen, F., in Proc. 17th International Workshop on
  Weak Interactions and Neutrinos (Cape Town, South Africa, January 1999)
  (astro-ph/9904216). 
\bibitem{fireball86}
  Paczy\'nski, B., Ap. J. {\bf 308}, L43 (1986);
  Goodman, J., Ap. J. {\bf 308}, L47 (1986).
\bibitem{internal} 
  Paczy\'nski, B. \& Xu, G., Ap. J. {\bf 427}, 708 (1994);
  M\'esz\'aros, P., \& Rees, M., MNRAS {\bf 269}, 41P (1994).
\bibitem{Woods95}
  Woods, E. \& Loeb, A., Ap. J. {\bf 453}, 583 (1995).
\bibitem{Bhat92}
  Bhat, P. N., {\it et al.}, Nature {\bf 359}, 217 (1992).
\bibitem{Gamma}
  Krolik, J. H. \& Pier, E. A., Ap. J. {\bf 373}, 277 (1991);
  Baring, M., Ap. J. {\bf 418}, 391 (1993).
\bibitem{BnM76}
  Blandford, R. D., \& Mckee, C. F., Phys. Fluids {\bf 19}, 1130 (1976).
\bibitem{Bland87} 
  Blandford, R., \& Eichler, D., Phys. Rep. {\bf 154}, 1 (1987).
\bibitem{Akerlof99}
  Akerlof, C. W. {\it et al.}, Nature, {\bf 398}, 400 (1999).
\bibitem{SPMR_0123}
  Sari, R. \& Piran, T., Ap. J. {\bf 517}, L109 (1999);
  M\'esz\'aros, P.  \& Rees, M., MNRAS {\bf 306}, L39 (1999).
\bibitem{GRB_z}
  Krumholtz, M., Thorsett, S. E., \& Harrison, F. A., Ap. J. {\bf 506}, L81
  (1998);
  Hogg, D. W. \& Fruchter, A. S., Ap. J. {\bf 520}, 54 (1999).
\bibitem{Freedman}
  Freedman, D. L., \& Waxman, E., submitted to Ap. J. (astro-ph/9912214).
\bibitem{MnM98plus}
  Mao, S. \& Mo, H. J., A\&A {\bf 339}, L1 (1998);
  Porciani, C., \& Madau, P., submitted to Ap. J. (astro-ph/0008294).
\bibitem{Bird934} 
  Bird, D. J., {\it et al.}, Ap. J. {\bf 424}, 491 (1994).
\bibitem{Yakutsk}
  Efimov, N. N. {\it et al.},  in {\it Proceedings of the International
  Symposium on Astrophysical Aspects
  of the Most Energetic Cosmic-Rays}, edited by M. Nagano and F. Takahara
  (World Scientific, Singapore, 1991), p. 20.
\bibitem{Takeda98} 
  Takeda, M. {\it et al.}, Phys. Rev. Lett. {\bf 81}, 1163 (1998).
\bibitem{composition}
  Gaisser, T. K. {\it et al.}, Phys. Rev. {\bf D47}, 1919 (1993);
  Dawson, B. R., Meyhandan, R., Simpson, K.M., Astropart. Phys. {\bf 9}, 
  331 (1998).
\bibitem{GZK}
  Greisen, K., Phys. Rev. Lett. {\bf 16}, 748 (1966); 
  Zatsepin, G. T., \& Kuzmin, V. A., JETP Lett., {\bf 4}, 78 (1966).
\bibitem{WnB_clustering}
  Waxman, E., \& Bahcall, J. N., Ap. J. in press (hep-ph/9912326)
\bibitem{RnM98} 
  Rachen, J. P., \& M\'esz\'aros,  P.,  Phys. Rev. {\bf D58},
  123005 (1998).
\bibitem{Woosley_Chevalier99}
  Woosley, S. E., \& MacFadyen, A. I., A\&AS, {\bf 138}, 499 (1999);
  Chevalier, R. A., \& Li, Z., Ap. J., {\bf 536}, 195 (2000).
\bibitem{Dai00}
  Dai, Z. G., \& Lu, T., submitted to Ap. J. (astro-ph/0002430).
\bibitem{HnH00}
  Alvarez-Muniz, J., Halzen, F., \& Hooper, D. W., Phys. Rev. D, in press
  (astro-ph/0006027).
\bibitem{John_book} 
  Bahcall, J. N., {\it Neutrino Astrophysics}, Cambridge 
  University Press (NY 1989).
\bibitem{atmo} 
  Fukuda, Y., {\it et al.}, Phys. Lett. {\bf B335}, 237 (1994);
  Casper, D., {\it et al.}, Phys. Rev. Lett. {\bf 66}, 2561 (1991); 
  Fogli, G. L., \& Lisi, E., Phys. Rev. {\bf D52}, 2775 (1995).


\end{thebibliography}
\end{document}